\documentclass[12pt]{article}
\usepackage{amsmath}
\input{epsf.sty}
\usepackage{hyperref}

\numberwithin{equation}{section}

\newcommand{\p}{\partial}
\newcommand{\e}{\epsilon}
\newcommand{\m}{\mu}
\newcommand{\n}{\nu}
\newcommand{\de}{\delta}

\newcommand{\bt}{\bar{T}}
\newcommand{\h}{\hat{t}_1}
\newcommand{\tht}{\hat{t}_2}
\newcommand{\ah}{\hat{A}}
\newcommand{\mh}{{\hat{\mu}}}
\newcommand{\nh}{{\hat{\nu}}}
\newcommand{\no}{\nonumber \\}

\def\href#1#2{#2}  

\begin{document}


\begin{titlepage}

\title{
\hfill\parbox{4cm}{
{\normalsize UT-Komaba/02-01}\\[-5mm]
{\normalsize\tt hep-th/0202079}
}
\\[40pt]
Realization of Brane Descent Relations \\
in Effective Theories
\\[30pt]
}
\author{
Koji {\sc Hashimoto}\thanks{Moved from ITP, USCB. 
{\tt koji@hep1.c.u-tokyo.ac.jp}}$\hspace{2mm}$
and
Satoshi {\sc Nagaoka}\thanks{
{\tt nagaoka@hep1.c.u-tokyo.ac.jp}}$\hspace{2mm}$
\\[20pt]
{\it Institute of Physics, University of Tokyo,}\\
{\it Komaba, Meguro-ku, Tokyo 153-8902, Japan}\\
}
\date{\normalsize February, 2002}    
\maketitle
\thispagestyle{empty}

\begin{abstract}
\normalsize\noindent 
We examine Sen's descent relations among (non-)BPS D-branes by using 
low energy effective field theories of D$p\bar{\rm D}p$ system. We find
that the fluctuation around the kink solution reproduces the low energy
matter content on a non-BPS D$(p-1)$-brane. The effective action for
these fluctuation modes turns out to be a generalization of
Minahan-Zwiebach model. In addition, it is shown that the fluctuations
around the vortex solution  consist of massless fields on a BPS
D$(p-2)$-brane and they are subject to Dirac-Born-Infeld action. We
find the universality that the above results do not refer to particular
forms of the effective action.
\end{abstract}

\end{titlepage}


\section{Introduction}

Sen's conjectures on tachyon condensation \cite{Sentac1} 
have contributed immensely to the development of study of
non-perturbative features in string theories. The conjectures are
basically composed of three parts: (a) The potential height of the
tachyon potential exactly cancels the tension of the original unstable
D-brane. (b) At the  stable true vacuum, the original D-brane disappears
and closed string theory is realized. (c) The kink-type tachyon
condensations correspond to lower dimensional D-branes and organize
``Descent relations'' between stable and unstable D-branes (see Fig.\
1). These conjectures have been examined from various viewpoints in 
string/M theory, such as boundary string field theories (BSFT) 
\cite{Witten1}, cubic/vacuum string field theory (C/VSFT), K-theory, 
and non-commutative solitons. The conjecture (a) was verified 
\cite{GS, Ghoshal, Andreev} in the BSFT, while the conjecture (b) 
has been challenged by the C/VSFT approach. The conjecture (c)
has been supported from the success of the K-theory, and the
non-commutative solitons gave a partial evidence also for (c). 

\begin{figure}[htb]
\begin{center}
  \begin{minipage}{10cm}
\begin{center}
\leavevmode
\epsfxsize=40mm
\epsfbox{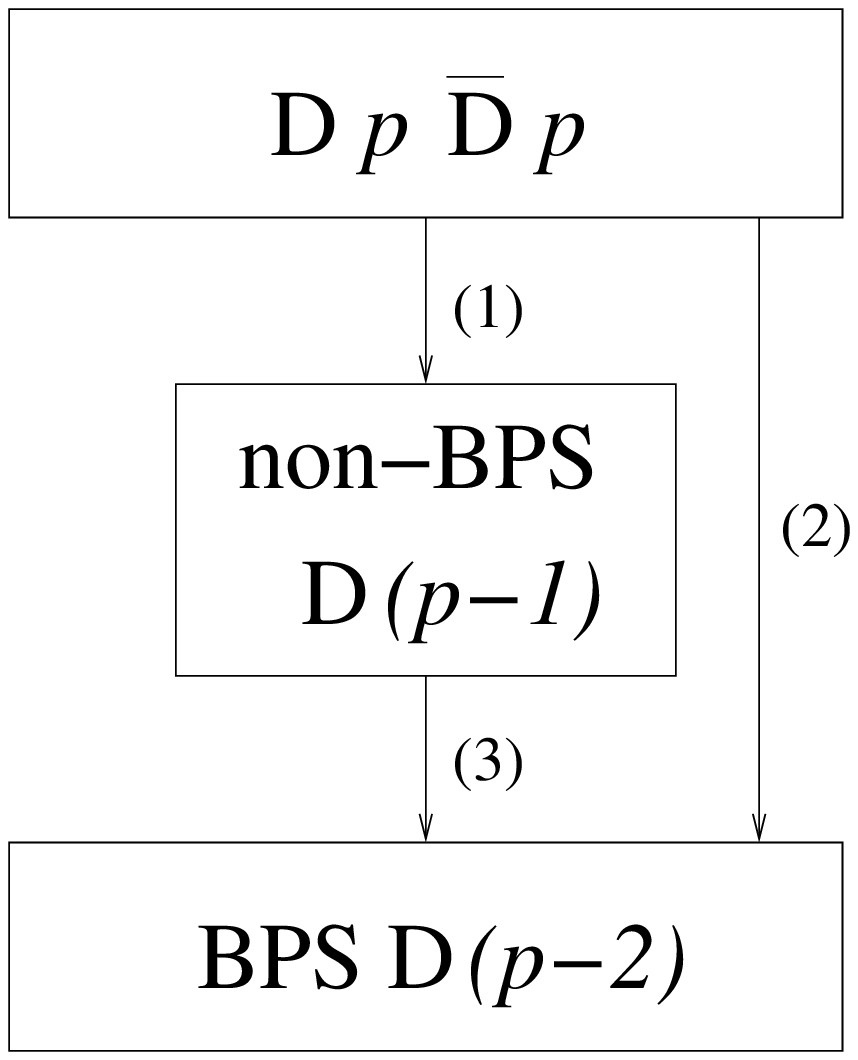}
\label{fig-dis}
\end{center}
  \end{minipage}
\hspace{-2cm}
\begin{minipage}{5cm}
Figure 1: The descent relations for the tachyon condensations. We
  omit the horizontal arrows which relate type IIA with type IIB,
  since they only denote the truncation of the fields.
\end{minipage}
\end{center}
\end{figure}

All of the above conjectures are involved with non-perturbative 
phenomena in string/M theory, thus non-perturbative analyses, such as
string field theories treating all the massive modes at once, are
necessary. However, we understand that the history of the development of
string theory in these six years tells us that the low energy effective
theory approach is powerful and promising. This approach is closely
related to the BSFT in which one turns on only certain profiles of
string excitation modes. Since the study of (b) requires
$s\leftrightarrow t$ channel duality, this needs all the massive modes
of open string excitation. However, the conjecture (c) is in our reach
even by the low energy effective field theory approach, if we
appropriately incorporate some of the BSFT results on the exact tachyon
potentials. 

The field theory models proposed by J.\ Minahan and B.\ Zwiebach (we
call MZ models) \cite{Zwie,MZ1,MZ2} gave a verification of a part of the
descent relations (the arrow (3) in Fig.\ 1 and the descent relations
among bosonic branes),  at the level of low energy
effective field theories (see also \cite{MN}). 
The MZ models include a tachyon field, and were proposed as a
toy model of the tachyon condensation. They have kink solutions
corresponding to co-dimension one BPS D-branes, and the fluctuations
around the solutions are exactly solvable. The resultant fluctuation
spectra have equal mass-squared spacing, which resembles to the spectra
of the BPS D-brane.  In fact, it turned out \cite{GS, MZ1, Andreev} 
that the MZ model is actually a two-derivative truncation of the BSFT
action of a non-BPS D-brane in type II string theories. This consistency
provided a further support to the verification by the MZ model.

In this paper, we investigate the descent relations (1) and (2) in
Fig.\ 1, by use of low energy effective field theories. Both arrows
have their starting point at the D$p\bar{\rm D}p$ system, therefore
we use effective actions for the D$p\bar{\rm D}p$ which are basically
derivative truncation of the BSFT results \cite{KL, Terashima}. 
The tachyon condensation (1) is the process of creating a 
co-dimension one defect, therefore we study a kink solution in the
D$p\bar{\rm D}p$ effective action. The kink represents the
non-BPS D($p-1$)-brane. Performing the fluctuation analysis around this
kink, we obtain an effective action describing the kink, and we show
that it coincides with the MZ model action which was the starting point
of the arrow (3) in \cite{MZ1, MZ2}. On the other hand, the process (2)
is the creation of a co-dimension two defect, a vortex, which represents
a BPS D($p-2$)-brane. 
Although Derrick's theorem prevents us from
obtaining naive co-dimension two solution, we consider higher derivative
terms and obtain a nonsingular vortex solution. We perform the
fluctuation spectra and obtain the massless spectra of the BPS brane,
and show that those modes obey Dirac-Born-Infeld (DBI) action which is a
correct low energy effective action of the BPS brane.
One should note that the arrow (2) is different from the naive
successive analyses (1) + (3), since the classical solution in (3) is
not exactly marginal in (1).

Furthermore, we can show the universality that our results do not
depend on the particular form of the action at the starting point. 
Requiring the properties followed from the BSFT, we write a rather
general action and reproduce most of the results above under a few
assumptions. We believe that our analysis provides a strong evidence for 
Sen's conjecture at the effective theory level.

The organization of this paper is as follows. In Sec.\ 2, after
reviewing the Minahan-Zwiebach model, we show the descent relation from
D9$\bar{\rm  {D}}$9 system to the non-BPS D8-brane in Type IIB string
theory concerning the tachyon condensation (1) by examining the
fluctuations around the kink solution of the D9$\bar{\rm  {D}}$9
effective action. In Sec.\ 3, we show the descent relation (2) which is
the tachyon condensation from the D9$\bar{\rm  {D}}$9 system to the BPS
D7-brane.  The universality in each case is shown in the respective
sections. We conclude with a summary and discussions in Sec.\ 4. The
appendix is concerning the construction of bound states of D-branes in
tachyon condensation. 


\section{Kink Solution}

In this section, we derive the Minahan-Zwiebach (MZ) model (which is 
a low-energy effective description of a non-BPS D8-brane) from a 
tachyon kink solution in a D9$\bar{\rm {D}}$9 system. This 
verifies a part of Sen's descent relation, the arrow (1) in Fig.\ 1, 
 at the level of effective field theories. 
We focus on the case $p=9$ for the rest of this paper,
and generalization to the other dimensions is straightforward.


\subsection{Review of Minahan-Zwiebach model} 

The Minahan-Zwiebach models \cite{Zwie,MZ1,MZ2} are low energy 
effective models which embody the tachyon dynamics for unstable 
D-branes in (super)string theories. Although the
models were first proposed as a toy model capturing desirable properties 
of string theories, it turned out \cite{GS, MZ1, Andreev} that one of
the models is a derivative truncation of the BSFT action of the  non-BPS
branes.   

Since we shall use some of the techniques developed by Minahan and
Zwiebach, in 
this subsection we review their methods. The ``superstring'' model which
is related to the BSFT is described by the action
\begin{eqnarray}
S={\cal T}\int\! d^{9}x\; e^{-T^2/a} 
\left(1 +(\p _\m T)^2\right) .
\label{MZaction}
\end{eqnarray}
This action is the two-derivative truncation of the tachyon action of
the BSFT.  Here ${\cal T}$ is the tension of the non-BPS D8-brane, and
the dimensionality of the numerical parameter $a$ is
retained\footnote{To obtain the precise mass squared for the tachyon
around the perturbative vacuum $T=0$, we have $a=2\alpha'$.} 
by $\alpha'$. The equation of motion is solved by a linear tachyon
profile $T=x_8$ which interpolates two vacua of the theory,
$T=\pm\infty$. The center of the kink sits at $x_8=0$, which can be
viewed from the fact that the perturbative unstable vacuum is at $T=0$. 
Let us consider the fluctuation $t$ around the kink solution,
\begin{eqnarray}
T=x_8+t(x) .
\end{eqnarray} 
Substituting this into the action (\ref{MZaction}), we have
\begin{eqnarray}
S= {\cal T}\int\! d^{9} x e^{-x_8^2/a}
\left(  \left(\frac{2x_8^2}{a^2}-\frac{1}{a}\right)2t^2
-\frac{4}{a}x_8t\p_8 t +(\p_\m t)^2
\right) .
\end{eqnarray}
After a partial integration and the field redefinition 
\begin{eqnarray}
\hat{t} \equiv e^{-x_8^2/2a} t,   
\label{redeft}
\end{eqnarray}
we obtain
\begin{eqnarray}
S={\cal T}\int d^{9}x \left( (\p_{\mh} \hat{t})^2+
\hat{t}\left(-\frac{\p^2}{\p x_8^2} +\frac{x_8^2}{a^2}
-\frac{1}{a}\right) 
\hat{t}
\right) .
\end{eqnarray}
where $\mh$ runs over all the directions except for $x_8$ which  
is the direction transverse to the brane. From the above expression,
the $x_8$ dependence of the fluctuation $\h$ obeys a Schr\"odinger
equation of a harmonic oscillator, thus the mass squared for the
fluctuation is equally spaced and specified by an integer $n$, 
\begin{eqnarray}
m^2_t=\frac{2n}{a} , \ n \geq 0 .
\end{eqnarray}
There is no tachyonic fluctuation, the mass tower starts from a massless
state and has the equal spacing. This result is consistent with the
identification of this kink as a BPS D7-brane. Furthermore, the mass
squared of the original tachyon field is $m_T^2=-\frac{1}{a}$, therefore
the spacing of mass levels is twice the value of $|m_T^2|$, as is also
consistent with string theory. 

If we include a gauge field in the original model, then we can get even
the gauge modes on the resultant kink BPS D7-brane. 
Observation in the BSFT tells us that the coupling between the gauge
field and the tachyon potential on the non-BPS D8-brane is provided 
through the Born-Infeld action as \cite{Andreev}
\begin{eqnarray}
S={\cal T}\int\! d^{9} x\; 
e^{-T^2/a} \sqrt{\det \left(1_{\m\n}+\sqrt{b}F_{\m\n}\right)} .
\end{eqnarray}
A sigma-model analysis shows that the parameter $b$ is
$(2\pi\alpha')^2$. We are working in the approximation of the
two-derivative truncation, 
\begin{eqnarray}
S={\cal T}\int \!d^{9} x\;
 e^{-T^2/a}\left(1+(\p_\m T)^2+\frac{b}{4} F^2
\right) .
\end{eqnarray}
The classical solution for the action is $T=x_8$, $A_\m =0$. The gauge
fluctuation around this solution is subject to the following second
order action 
\begin{eqnarray}
S=b {\cal T}
\int\! d^{9} x\;  e^{-T^2/a} \left( \frac{1}{4} F_{\mh \nh} F^{\mh \nh}
+\frac{1}{2} \p_8 A_{\mh} \p_8 A^{\mh}
\right) ,
\label{acga}
\end{eqnarray}
where we have chosen the axial gauge condition $A_8 =0$. 
After the field redefinition $\ah_{\mh}\equiv e^{-x_8^2/2a} \ah_{\mh}$,
the action (\ref{acga}) is rewritten as 
\begin{eqnarray}
S=b{\cal T} 
\int\! d^{9} x \left[\frac{1}{4} \hat{F}_{\mh\nh} \hat{F}^{\mh\nh}
+\frac{1}{2} \ah_{\mh}
\left(-\frac{\p^2}{\p x_8^2}+\frac{x_8^2}{a^2}-\frac{1}{a}\right)
\ah_{\mh}
\right] .
\end{eqnarray}
Therefore, the mass spectrum of the gauge fluctuation starts from
massless gauge modes and is equally spaced,
\begin{eqnarray}
m^2_{A_{\mh}} =\frac{2n}{a} , \quad  n \geq 0 .
\end{eqnarray}
This result is also consistent with the expectation from the spectra
on the BPS D7-brane. 

The Minahan-Zwiebach ``superstring'' model 
provides a suitable example realizing 
various aspects of the tachyon condensation on non-BPS branes,
in particular a part of the descent relations.\footnote{Furthermore, it
  is known that the massless fields obtained here generally obeys the
  dimensionally reduced Born-Infeld action which is an effective action
  of a D7-brane \cite{Tseytlin2}. } 
The background consistency with the BSFT ensures this 
success, and we employ this philosophy to check the 
other part of the descent relations in the rest of this paper.


\subsection{Kink solution in brane antibrane system}

The other important part in Sen's descent relations is the relation
between the D9$\bar{\rm {D}}9$ and a non-BPS D8-brane.
The latter is conjectured to be a kink
solution of the former. In order to show this conjecture at the
effective theory level, first we follow the procedures
of the analysis reviewed in the previous subsection.

The effective action of the D9$\bar{\rm {D}}9$ system in the two
derivative truncation is written as \cite{KL,Terashima}
\begin{eqnarray} \label{98-ac}
S={2T_{\rm  D9}} \int d^{10} x \ e^{-|T|^2/a} 
\Big(1+|D_\m^{(-)} T|^2+\frac{1}{4}b (F_{\m \n}^{(-)})^2
+\frac{1}{4}b (F_{ \mu \nu }^{(+)} )^2
\Big) ,
\end{eqnarray}
where $T\equiv T_1 + i T_2$ is a complex tachyon charged under 
the gauge field $A_\m^{(-)}$,
\begin{eqnarray}
D_\m^{(-)}T=\p_\m T-iA_\m^{(-)}T.
\end{eqnarray}
The fields $A_\m^{(\pm)}$ are the 
linear combinations of the gauge fields on the brane
and the antibrane,
\begin{eqnarray}
A_\m^{(\pm)}=A_\m ^{(1)}\pm A_\m^{(2)} ,
\end{eqnarray}
and $a,b$ are dimensionful numerical constant which can be
determined by the BSFT analysis, string sigma model calculation, or
string scattering amplitude.\footnote{To be precise, 
here we have $a=2\alpha'$ and $b=(2\pi\alpha')^2$, as mentioned in the
previous subsection. Note that the value of $a$ depends on the
renormalization scheme in the BSFT and the sigma model approach
\cite{Tseytlin1}. 
} 

\vspace*{5mm}
\noindent
\underline{Classical solution }

The equations of motion for this action are
\begin{eqnarray} \label{98-eq}
&&-\frac{1}{a} 
\left[
\bt\left(1+\frac{b}{4} (F_{\m \n}^{(-)})^2
+\frac{b}{4} (F_{ \mu \nu }^{(+)} )^2
\right) - T (D_\m \bt)^2
\right]
-D_\mu D_\mu \bt =0 ,
\\
&&
\p_\n (e^{-|T|^2/a} F_{\m\n}^{(+)})=0 ,
\\
&&
ie^{-|T|^2/a}(TD_\m \bt - \bt D_\m T) +b\p_\n(e^{-|T|^2/a}
F_{\m\n}^{(-)})=0 .
\label{third}
\end{eqnarray}
Note that putting $T_2=A^{(-)}=0$ makes the action and equations of
motion reduce to those of the MZ model. In particular, the
third equation (\ref{third}) 
which came from the differentiation with respect to
$A^{(-)}$ is trivially satisfied by this condition.
Therefore, the kink solution of the MZ model is inherited to our
situation. The desired solution is 
\begin{eqnarray}
T=x_9,
\quad
A_\mu^{(\pm)}=0.
\label{98-sol}
\end{eqnarray}

\vspace*{5mm}
\noindent
\underline{Fluctuation analysis}
\vspace*{2mm}

Let us consider the fluctuation around this kink solution.
As is obvious, the fluctuation spectra for the fields $T_1$ and
$A_\mu^{(+)}$ are precisely the same as in the previous subsection,
thus we obtain 
\begin{eqnarray}
m^2_{t_1}=\frac{2n_1}{a} , \quad 
m^2_{A_\mh^{(+)}}= \frac{2n_2}{a}  \qquad 
\mbox{where}\quad \ n_i \geq 0, \quad i=1,2.
\label{98-fluta}
\end{eqnarray}
Here we denote the fluctuation of $T_1$ as $t_1$. 
The lowest modes of these fluctuations are zero modes,
and especially the $t_1$ zero mode corresponds to the
scalar field on the non-BPS D8-brane, representing the transverse
displacement of the brane. 
This must have appeared  by the Nambu-Goldstone theorem, 
since the translational invariance along $x_9$
direction is broken by the kink solution.

Another notable point about the zero modes 
is the $x_9$ dependence. Since we solved the harmonic oscillator
problem for $\hat{t}_1$ and $\ah_\m^{(+)}$, 
the ground state's wave functions 
are gaussian:
\begin{eqnarray}
\mbox{[$x_9$ part of $\h (x_0,x_1,\cdots,x_9)$]} = 
e^{-\frac{x_9^2}{a}}.
\end{eqnarray}
Therefore, turning back to the field redefinition (\ref{redeft}), 
we can show that the zero modes of $t_1, A_\mu^{(+)}$ 
are independent of $x_9$. This kind of the property of the lowest
fluctuation was observed first in the bosonic D-brane system
\cite{MN}. 

The fluctuation analysis for the other fields $T_2$ and $A_\mu^{(-)}$
can be performed in the same manner.\footnote{We treat fluctuations
of the fields one by one separately, since generically it is difficult 
to diagonalize all of the fluctuations. We shall discuss this point in
the final section.} 
First, we consider the fluctuation of $T_2$ around the 
classical solution which is written as 
\begin{eqnarray}
T=x_9 +i t_2 (x), \ A_\m^{(\pm)}=0 .
\end{eqnarray}
Substituting this into the action (\ref{98-ac}) and picking up the
second order terms, we obtain
\begin{eqnarray}
S_{t_2}= {2T_{\rm  D9}} \int d^{10} x \
e^{-\frac{x_9^2}{a}}
\left(-\frac{2}{a}t_2^2 +(\p_\m t_2)^2
\right) . 
\end{eqnarray}
After the same redefinition ${\tht } =e^{-\frac{x_9^2}{2a}} t_2$ ,
the fluctuation action is written as 
\begin{eqnarray}
S={2T_{\rm  D9}} \int d^{10} x 
\left(
\left(\frac{x_9^2}{a^2} -\frac{3}{a}\right){\tht}^2+(\p_\m \tht )^2 
\right) .
\end{eqnarray}
Therefore the series of the mass squared begins with 
the negative mass squared, which is expected to exist
as a tachyonic mode of the non-BPS D8-brane, 
\begin{eqnarray}
m_{t_2}^2=\frac{2(n-1)}{a} , \quad n \geq 0 .
\label{98-t2}
\end{eqnarray}
The mass squared spacing is the same as that of $t_1$ fluctuations.
Again, the lowest mode is independent of $x_9$. 

Since there should be a single massless gauge field on the resultant
non-BPS D8-brane, the fluctuation of another gauge field $A_\m^{(-)}$ 
must be massive. We shall see this below. 
The fluctuation we consider here is 
\begin{eqnarray}
T=x_9,\ A_\m^{(+)}=0,\ A_\m^{(-)}=A_\m^{(-)} (x) .
\end{eqnarray}
Following the same procedures with the field redefinition 
$
{\ah_\m^{(-)}} =e^{-\frac{x_9^2}{2a}} A_\m^{(-)} ,
$
we have
\begin{eqnarray}
S={2T_{\rm  D9}} \int d^{10} x 
\left( \frac{b}{4} \left(\hat{F}_{\mh\nh}^{(-)}\right)^2
+ \frac{b}{2}\ah_{\mh}^{(-)} \left[-\p_9^2 +(\frac{2}{b}
+\frac{1}{a^2})x_9^2-\frac{1}{a}\right]
\ah_{\mh}^{(-)}
\right) ,
\end{eqnarray}
where the indices $\mh,\nh$ run over $0,1,\cdots,8$,
and the gauge fixing condition $\p_{\mh} {\ah}_{\mh}^{(-)}=0$ has been used.
The mass squared is given as
\begin{eqnarray}
m^2_{A_\mh^{(-)}}= 2n\sqrt{\frac{2}{b}+\frac{1}{a^2}}
+\sqrt{\frac{2}{b}+\frac{1}{a^2}}-\frac{1}{a} , \quad n \geq 0.
\label{fluctuationA-}
\end{eqnarray}
The lowest mode of these fluctuations is massive, 
as expected. It does not appear in
the low energy effective theory of the kink.

We have seen that $A^{(+)}$ still has a massless mode while $A^{(-)}$
becomes massive. One of the physical explanation is as follows. The
tachyon field couples only to $A^{(-)}$, thus if the tachyon condenses,
the gauge field $A^{(-)}$ is Higgsed and becomes
massive.\footnote{However, our tachyon condensation is a linear profile,
hence a naive mass of the Higgsed gauge field $A^{(-)}$ vanishes at
$x^9=0$. This is against our expectation and too naive, thus we have
needed the precise fluctuation analysis (\ref{fluctuationA-}).}
On the other hand, $T$ is neutral under $A^{(+)}$, thus $A^{(+)}$ stays
massless after the tachyon condensation. One can see this in another
way. The gauge field $A^{(+)}$ is responsible for the lower-dimensional
D-brane charges, as seen in App.\ A as an example. Therefore $A^{(+)}$
should remain massless on the kink (or also on the vortex as we shall
see in the next section).

\vspace*{5mm}
\noindent
\underline{Effective action}

In the end, we summarize this subsection by writing the total
effective action obtained from the above fluctuation analysis.
The gauge field $A^{(-)}$ does not appear in the low energy region.
We substitute the lowest fluctuation modes into the action (\ref{98-ac}) 
and obtain  
\begin{eqnarray}
S_{\rm {tot}}\!\!\!
&& = {2T_{\rm  D9}} \int d^{10} x 
e^{-(x_9+t_1)^2/a-t_2^2/a}
\left(
2+(\p_{\mh} t_1)^2 +(\p_{\mh} t_2)^2 
+\frac{b}{4} (F^{(+)}_{\mh\nh})^2
\right) 
\no
&&=
{2T_{\rm  D9}} \int\! d^{9} x \;
e^{-t_2^2/a}\left(
2+(\p_{\mh} t_1)^2 +(\p_{\mh} t_2)^2 
+\frac{b}{4} (F^{(+)}_{\mh\nh})^2
\right) 
\int d(x_9+t_1(x_\mh)) e^{-(x_9+t_1)^2/a}
\no
&&=
{2T_{\rm  D9}} \sqrt{a\pi}
\int\! d^{9} x \;
e^{-t_2^2/a}\left(
2+(\p_{\mh} t_2)^2 +(\p_{\mh} t_1)^2 
+\frac{b}{4} (F^{(+)}_{\mh\nh})^2
\right) .
\label{effec}
\end{eqnarray}
In the second equality, we have used the fact that 
$t_1,t_2,$ and $A_\mu^{(+)}$ are independent of $x_9$.
The third equality is obtained by the integration over the transverse
direction. Due to the gaussian factor, we can say that the 
defect described by the action (\ref{effec}) is located
at $x_9 = -t_1(x_\mh)$. 
This action (\ref{effec}) 
is precisely identical with the effective action of a non-BPS D8-brane.
The difference from the action of the MZ model (\ref{MZaction}) is that
now it is generalized to incorporate the transverse 
scalar field on the non-BPS brane.  The way it is included in is just
the same as the two-derivative truncation of the BSFT action for the 
non-BPS D8-brane. 

Thus, we have shown the decent relation from D9$\bar{\rm {D}}$9
system to the non-BPS D8-brane by directly examining the fluctuations
around the kink solution of the D9$\bar{\rm {D}}$9 effective action.


\subsection{General action and universality of the results
  \label{gen}}

Although the results obtained in the previous subsection are desirable,
they may depend on the choice of the effective theories. 
In this subsection, we show that the results of the previous
subsection are in fact quite general, in the sense that the results do
not depend on the specific form of the Lagrangian. 
If we remember that the preferable properties of the MZ model are
intrinsically due to its relation to the BSFT, the general Lagrangian
should be constrained by some results of the BSFT. Here we respect the 
properties of the BSFT action derived for linear tachyon profiles. 
Since in the BSFT it is difficult to treat $A^{(-)}$ generically, we
neglect this gauge field in this general approach.

The symmetries which the model should have are (i) Lorentz invariance,
(ii) gauge invariance with respect to $A^{(+)}$, 
and (iii) the global gauge invariance associated originally with
$A^{(-)}$, 
\begin{eqnarray}
  T \mapsto e^{i\theta} T.
\label{globalgauge}
\end{eqnarray}
Employing the BSFT and the sigma-model result, the coupling between
the gauge field $A^{(+)}$ and the tachyon is determined 
(under the assumption that we
consider the constant gauge field strength and neglect the tachyon
higher derivatives $\p\p T$) as  
\begin{eqnarray}
S={2T_{\rm  D9}} \int d^{10} x \ e^{-|T|^2/a} \sqrt{\det
  \left(1+\sqrt{b}F\right)}
f(X,Y).
\label{geneac}
\end{eqnarray}
In the above expression the invariants $X$ and $Y$ are defined as
\begin{eqnarray}
X \equiv G^{\m\n}\p_\m T \p_\n \bt, 
\quad Y \equiv |G^{\m\n}\p_\m T\p_\n T|^2,
\end{eqnarray}
and the metric $G$ is in the well-known form, 
\begin{eqnarray}
G^{\m\n} =\left(\frac{1}{1-bF^2}\right)^{\m\n} .
\end{eqnarray}
We omit the upper case $(+)$ here and in the rest of this subsection.

\vspace*{5mm}
\noindent
\underline{Classical solution}

In order for the fluctuation analysis to be performed, 
we require that a classical solution of the above system 
should be a linear profile of the tachyon:
\begin{eqnarray}
T=qx_9,\quad A_\m=0.
\end{eqnarray}
Then the equations of motion become a constraint on
the unknown function $f(X,Y)$: 
\begin{eqnarray} \label{sft-eq3}
\left[f-2q^2\frac{{\de} f}
{{\de} X}-4q^4\frac{{\de} f}{{\de} Y}
\right]_{X=q^2, Y=q^4}=0 .
\end{eqnarray}

\vspace*{5mm}
\noindent
\underline{Fluctuation analysis}
\vspace*{2mm}

First let us carry out the $T_1$ fluctuation. 
Substituting $T=qx_9+t_1(x)$ and $A_\m=0$ into the action
(\ref{geneac}) and picking up the quadratic terms, we obtain
\begin{eqnarray}
S={2T_{\rm  D9}} \int d^{10} x \ 
e^{-q^2x_9^2/a}
\left(C_1(\p_8 t_1)^2+\frac{f}{2q^2}(\p_{\mh} t_1)^2
\right) ,
\end{eqnarray}
where
\begin{eqnarray}
C_1 \equiv 
\left[
\frac{{\de} f}{{\de} X}
+6q^2\frac{{\de} f}{{\de} Y}
+2q^2\frac{{\de}^2 f}{{\de} X^2}
+8q^6\frac{{\de}^2 f}{{\de} Y^2}
+8q^4\frac{{\de}^2 f}{{\de} X {\de} Y} 
\right]_{X=q^2, Y=q^4}.
\end{eqnarray}
Note that we use the equation of motion (\ref{sft-eq3}) here.
Then the mass squared is
\begin{eqnarray}
m_{t_1}^2=\frac{4q^4C_1}{af(q^2,q^4)} n , \quad n \geq 0 ,
\end{eqnarray}
therefore there is a massless mode which corresponds to the 
transverse scalar field on the non-BPS D8-brane,
that is the same as in the derivative truncated case.

It is straightforward to check that this is also the case for the 
$T_2$ fluctuation.
Substituting $T=qx_9+it_2(x), A_\m=0$ into the action, we have
\begin{eqnarray}
S={2T_{\rm  D9}} \int\! d^{10} x
\left(
\frac{f(q^2,q^4)}{2q^2} \tht
\left(
-\!\p_9^2\!+\!\frac{q^4}{a^2} x_9^2\!-\!\frac{3q^2}{a}
\right){\tht}
+
(\p_\m \tht)^2
\left[
\frac{\de f}{\de X}\!-\!2q^2\frac{\de f}{\de Y}
\right]_{\!\!
  \begin{array}{l}
\mbox{\scriptsize $X=q^2$} \\[-2mm]
 \mbox{\scriptsize $Y=q^4$}
  \end{array}
} \!\!
\right) .
\nonumber
\end{eqnarray}
The mass squared is
\begin{eqnarray}
m_{t_2}^2=
(n-1)
\left[
\frac{f}{a}
\left(\frac{\de f}{\de X}-2q^2 \frac{\de f}{\de Y}\right)^{-1}
\right]_{ \!\!\!\!
  \begin{array}{l}
\mbox{\scriptsize $X=q^2$} \\[-2mm]
 \mbox{\scriptsize $Y=q^4$}
  \end{array}
}  , \quad n\geq 0 ,
\end{eqnarray}
thus there is a tachyonic mode. 

However, the result for the gauge fluctuation is a little bit
different. The analysis for the gauge fluctuations gives the following
quadratic  action,
\begin{eqnarray}
S={2T_{\rm  D9}}
\frac{bf(q^2,q^4)}{4} 
 \int d^{10} x \ 
e^{-\frac{q^2}{a}x_9^2} 
F_{\mh\nh}^2 .
\label{gaugeflu}
\end{eqnarray}
The remarkable point is the vanishing of the $A_{\mh}^2$ terms, 
therefore we have no massive tower, as opposed to the case of the
two-derivative truncation. We have only the massless gauge mode
localized on the kink solution. Note that, as for the gauge field, 
our general analysis does not include the action (\ref{98-ac}) 
of the previous subsection, because we have another contribution for
$F^2$ from 
the metric $G^{\m\n}$. 

The lowest modes for the tachyon fluctuations $t_1$ and $t_2$ 
are independent of $x_9$, as before. 
On the other hand, as seen above, the gauge fluctuation has no tower 
and hence no information on its $x_9$ dependence, thus we cannot include
the gauge fluctuation in obtaining the effective action of the kink
in the same manner as (\ref{effec}). However, in the expression 
(\ref{gaugeflu}) the gaussian factor in front of the 
field strength behave as a space-dependent dielectric constant
\cite{Kuroki}, thus the massless gauge field is localized on the kink. 
We conclude that we have obtained the precise field content on the
non-BPS D8-brane by the fluctuation analysis in the general action 
(\ref{geneac}).


\section{Vortex Solution}

As mentioned in the introduction, how the stable BPS branes appear in
the tachyon condensation of the brane antibrane system is the basis
of the K-theory argument. After the development of the techniques to
deal with the tachyon condensation such as the BSFT, we can now closely
look at the nucleation of the lower dimensional BPS D-branes after the
tachyon condenses. In this section, we focus on the creation of a BPS 
D7-brane as a vortex solution,\footnote{We use the term ``vortex'' in a
broad sense, not only for the usual gauge vortex. Following Sen's
original usage, we specifically call the tachyon configuration
(\ref{4d-sol}) ``vortex''. } from the tachyon condensation of the
D9$\bar{\rm {D}}$9 system. After constructing the solution, we
perform the fluctuation analysis and then see if the precise
worldvolume field content appears on the vortex D7-brane. 
We will see that the general analysis independent of the form of the
action, studied in Sec.\ 2.3, can be applied also here. Finally we
derive the DBI action as an effective field theory of the fluctuation
modes. 

\subsection{Vortex solution for the tachyon field}

Since the inclusion of the gauge fields is accompanied by some
complication which we shall see in Sec.\ 3.3, first we 
consider a simple situation where we treat only the complex 
tachyon field. The action in the two derivative truncation is 
written as
\begin{eqnarray} 
S={2T_{\rm  D9}} \int d^{10} x \ e^{-|T|^2/a}
\left(1+  |\p_\m T|^2 
\right) .
\label{trun2}
\end{eqnarray}
Although the D7-brane should be realized as a co-dimension two vortex
solution of the tachyon system, the well-known Derrick's theorem
\cite{Derrick} prohibit us from proceeding. 
Let us recall how Derrick's theorem applies to our situation with
the action (\ref{trun2}). 
Make the tachyon profile to depend only on two coordinates 
among the worldvolume directions, and suppose that we have a finite
energy solution $T=T_{\rm {sol}}$. 
Then, the rescaling 
$T_{\rm {sol}}(x) \mapsto T' =T_{\rm {sol}}(\lambda x)$ 
changes the energy (action) to
\begin{eqnarray}
S'&&= {2T_{\rm  D9}} V_{\rm D7}
  \int d^{2} x  \ e^{-|T'|^2/a}
  \left(1+  |\p_\m T'|^2 \right) \no
&&= {2T_{\rm  D9}} V_{\rm D7} 
   \int d^{2} x'  \ e^{-|T_{\rm {sol}}(x')|^2/a}
  \left(\lambda^2 + |\p'_\m T_{\rm {sol}}(x')|^2 \right) ,
\end{eqnarray}
where $x'\equiv\lambda x$. Since $T_{\rm {sol}}$ with $\lambda=1$ 
should minimizes the energy  
$\frac{d}{d\lambda}E(T_{\rm {sol}})=0$, we obtain
\begin{eqnarray}
\int d^{2} x  \ e^{-|T_{\rm {sol}}|^2/a}=0 ,
\end{eqnarray}
which tells us that the tachyon solution should sit at the potential
minimum everywhere, $|T_{\rm {sol}}|=\infty.$
This shows that there exists no vortex solution as far as 
we employ the action (\ref{trun2}). 
To avoid this no-go theorem, we have to include terms supplying
negative powers in $\lambda$ to stabilize the system against the
rescaling, such as 
tachyon higher derivative terms or gauge field kinetic
terms. In this and the next subsections, we adopt the first approach.
For the latter approach, see \cite{Suyama}.\footnote{The action
used in \cite{Suyama} is slightly different from the BSFT results.} 

Let us consider an action of the four derivative truncation. 
The form of the additional terms respects the BSFT
results\cite{KL,Terashima}, 
\begin{eqnarray} \label{4d-ac}
S={2T_{\rm  D9}} \int d^{10} x \ 
e^{-|T|^2/a}
\left(1+  |\p_\m T|^2 +p_1 |\p_\m T |^4
 +p_2 (\p_\m T)^2(\p_\n \bt)^2
\right).
\end{eqnarray}
New coefficients $p_1$ and $p_2$ are some numerical constant.
The origin of these higher derivative terms is as follows. 
The BSFT action obtained in \cite{KL,Terashima} 
for the linear tachyon profile is given by
\begin{eqnarray}
S=\int d^{10} x \; e^{-|T|^2} \Pi_{i=1,2} F(u_i^2) ,
\end{eqnarray}
where $T=u_1 x_8 +iu_2 x_9$ and $F(x)=\frac{4^x x
  \Gamma(x)}{2\Gamma(x)}$.
We omit numerical coefficients and $\alpha'$ for simplicity. 
Performing Taylor-expansion for this action around $u=0$ and 
restoring the Lorentz invariance
and the global gauge invariance (\ref{globalgauge}), 
we obtain the action (\ref{4d-ac}). 

We can rephrase
this in another way. Suppose that the action can be
factorized  as 
\begin{eqnarray}
  S={2T_{\rm  D9}}\!\int \!d^{10}x \;
e^{-|T|^2/a} f(\p T), 
\end{eqnarray}
then the general terms of the order $(\p T)^4$ in $f$, which are Lorentz
invariant and global gauge invariant, are only the ones in
the action (\ref{4d-ac}). 

\vspace*{3mm}
\noindent
\underline{Classical solution} 

Assuming that the tachyon solution is linear, 
the equations of motion for this action (\ref{4d-ac}) reduce to
\begin{eqnarray} \label{4d-eom}
\bt
\left[
1-p_1|\p_\mu T|^2 - p_2 (\p_\mu T)^2(\p_\nu \bt)^2
\right]
-T (\p_\mu \bt)^2
\left[
1+2(p_1 + p_2)|\p_\nu T|^2
\right]
=0.
\end{eqnarray}
A linear vortex solution is found as
\begin{eqnarray}
&T=q(x_8+ix_9), \label{4d-sol}
\end{eqnarray}
where the real parameter $q$ satisfies $4 q^4p_1=1$. 
The true vacuum of the tachyon potential $e^{-|T|^2/a}$ is $S^1$ at
the infinity of the tachyon configuration space, $|T|=\infty$. We deal
only with the co-dimension two solitons, hence the solitons are
categorized by the homotopy group $\pi_1(S^1)={\sf
  Z\hspace{-5pt}Z}$ characterizing the winding number of the
solitons. The above solution (\ref{4d-sol}) has a unit 
charge in this sense. 

The reason why we have assumed the solution to be linear
is that linear solutions with the winding number one 
are directly related to the Ramond-Ramond charge
of the BPS D7-brane, through the Chern-Simons (CS) term
of the D9$\bar{\rm {D}}$9 effective action computed in
\cite{KL,Terashima}. 
The relevant CS term is given by
\begin{eqnarray} \label{4d-CS} 
S_{\rm {CS}} 
={2T_{\rm  D9}}  \int C \wedge {\rm  Tr}\sigma_3 \exp 
 \left(\begin{array}{cc}
      -|T|^2  &  d\bt \\
      dT &  -|T|^2
        \end{array}\right) ,
\end{eqnarray}
where we set the gauge field to vanish, and omit numerical parameters
for simplicity.
Substituting the solution (\ref{4d-sol}) into the above expression
(\ref{4d-CS}), we can obtain the D7-brane charge,
\begin{eqnarray}
S_{\rm {CS}} ={2T_{\rm  D9}} \int e^{-|T|^2}
C^{(8)} \wedge
dT \wedge d\bt
\sim  \int_{V_{\rm {D7}}} C^{(8)} .
\end{eqnarray}

\vspace*{5mm}
\noindent
\underline{Fluctuation analysis} 
\vspace*{2mm}

Since the solution has a symmetry of 
interchanging $x_8$ and $x_9$ simultaneously
with the exchange of $T_1$ and $T_2$, it is sufficient to consider the
fluctuation of $T_1$ around the vortex solution (\ref{4d-sol}), 
$T=q (x_8+ix_9) + t_1(x)$.
Substituting this into (\ref{4d-ac}), we have the action
\begin{eqnarray} 
S\!\!\!\!&&={2T_{\rm  D9}} \int d^{10} x \ 
e^{-\frac{q^2(x_8^2+x_9^2)}{a}}
\left[
2\left(\frac{2q^2}{a^2}x_8^2  -\frac{1}{a}\right)(1+q^2)t_1^2
-\frac{4(1+q^2)}{a} x_8 t_1 \p_8 t_1 
\right.\no
&& \hspace*{2cm}
\left.
+\frac{1+q^2}{q^2}(\p_\mh t_1)^2 +\frac{1}{q^2}(\p_8 t_1)^2 
+4p_2 q^2\big( (\p_8 t_1)^2+(\p_9 t_1)^2 \big)
\right] ,
\end{eqnarray}
where we have extracted only the terms quadratic in $t_1$.
After performing the field redefinition 
\begin{eqnarray}
\h =t_1\exp\left[-\frac{q^2(x_8^2+x_9^2)}{2a}\right]  ,
\end{eqnarray}
we can eliminate the exponential factor in the action as
\begin{eqnarray}
S=\!\!\!\!&&
 {2T_{\rm  D9}} \int d^{10} x \ 
\left(
(2+q^2+4p_2q^4)
\left(
\frac{q^2}{a^2} x_8^2 \h^2
+\frac{1}{q^2}(\p_8 \h)^2 -\frac1a \h^2
\right)
\right.
\no
&&
\left.
+ (1+q^2+4p_2q^4)
\left(
\frac{q^2}{a^2} x_9^2\h^2 
+ \frac{1}{q^2}(\p_9 \h )^2- \frac{1}{a} \h^2
\right)
+\frac{1+q^2}{q^2}(\p_{\mh} \h)^2
\right)
\end{eqnarray}
where $\mh$ denotes the directions along the worldvolume of the
defect, $0,1,\cdots ,7$. Solving the two-dimensional harmonic
oscillator problem, the mass squared is found as
\begin{eqnarray}
m^2_{t_1}=\frac{2q^2}{a(1+q^2)}
\Big(
(2+q^2+4p_2q^4)n_1 +  (1+q^2+4p_2q^4) n_2
\Big), \quad
n_1, n_2 \geq 0.
\label{97-t1}
\end{eqnarray}
The lowest mode for the fluctuations is massless and corresponds to
the Nambu-Goldstone (NG) boson associated with the breaking of the
translational invariance along $x_8$. This mode is in fact 
the transverse scalar field on the BPS D7-brane.
When we replace $t_1$ by $t_2$ which is the fluctuation for $T_2$, 
from the symmetry mentioned above, we obtain the same result and 
get the transverse NG mode for the $x_9$ direction. This is also
expected to exist on the D7-brane.


\subsection{Vortex solution in the brane antibrane system with gauge fields}

Let us proceed to include the gauge fields on the D9$\bar{\rm {D}}$9 system.
First we treat the gauge field $A_\m^{(+)}$.
We employ a similar gauge coupling as the MZ model, then the total action
is written as
\begin{eqnarray} \label{ap-ac}
S={2T_{\rm  D9}} \int d^{10} x \ 
e^{-|T|^2/a}
\left(1+  |\p_\m T|^2 +\frac{b}{4}( F^{(+)})^2+
p_1 
\left(|\p_\m T|^2\right)^2
+p_2 (\p_\m T)^2(\p_\n \bt)^2
\right) .
\end{eqnarray}
A classical solution naively extended is
\begin{eqnarray}
T=q(x_8+ix_9),\quad
A_\m^{(+)}=0 ,
\end{eqnarray}
and we consider the fluctuation $A_\m^{(+)}=A_\m (x)$ around this
solution.
Substituting these into (\ref{ap-ac}), we obtain
\begin{eqnarray}
S=&&{2T_{\rm  D9}} \int\! d^{10} x \; 
\left(
\ah_{\mh}
  \left[
    -\p_8^2-\p_9^2+\frac{q^4}{a^2}(x_8^2+x_9^2)-\frac{2q^2}{a}
  \right] 
\ah_{\mh}
+\frac{1}{2}\hat{F}_{\mh \nh}^2
\right.
\no
&&
+
  \ah_8\left[
    - \p_{\mh}^2 -\p_9 +\frac{q^4}{a^2}x_9^2-\frac{q^2}{a}
  \right] \ah_8
  +\ah_9\left[
     -\p_{\mh}^2 -\p_8^2+\frac{q^4}{a^2}x_8^2-\frac{q^2}{a}
  \right] \ah_9 
\no 
&&
+\frac{2q^2}{a}x_8\ah_8[\p_{\mh}\ah_{\mh}+\p_9 A_9] 
+\frac{2q^2}{a}x_9\ah_9[\p_{\mh}\ah_{\mh}+\p_8 A_8]
-\frac{2q^4}{a^2}x_8x_9\ah_8 \ah_9 
\no 
&& \hspace*{2cm}
-2(\p_8 \ah_8 \p_9 \ah_9+\p_8 \ah_8 \p_{\mh} \ah_{\mh}+\p_9 \ah_9\p_{\mh} 
\ah_{\mh})
\Big) ,
\end{eqnarray}
where we have already redefined the gauge field as
$\ah_\m\equiv  A_\m\exp(-q^2(x_8^2+x_9^2)/2a)$.
This action has cross terms between $\hat{A}_\mh$ and $\hat{A}_{8,9}$,
hence it is difficult to diagonalize this action. But if we choose the
gauge condition $\p_\mh \ah_{\mh}=0$,
the $\hat{A}_\mh$ part can be separated from the rest
$\hat{A}_{8.9}$. In the following, we shall consider only the separated
$\hat{A}_\mh$ part which is the first line of the above action. 
The mass squared is easily obtained as 
\begin{eqnarray}
m_{A_{\mh}^{(+)}}^2=\frac{2q^2n}{a}, \quad n \geq 0 .
\label{97-a+}
\end{eqnarray}
Thus, $A_{\mh}^{(+)}$ has a single massless gauge field on the vortex, 
which is one of the correct massless content on the BPS D7-brane.

Second, we move to the analysis of the fluctuations for $A_\m^{(-)}$.
The action we should consider here is the following: 
\begin{eqnarray}
S=&&\!\!\!\!\!{2T_{\rm  D9}} \int d^{10} x \ 
e^{-|T|^2/a}
\Big(1+  D_\m T D^\m \bt +
\frac{b}{4} (F^{(-)})^2+ i c F^{(-)}_{\m\n}D_\m TD_\n \bt
\no &&\hspace*{5cm} 
+p_1  (D_\m T D^\m \bt)^2
+p_2 (D_\m T)^2(D_\m \bt)^2
\Big) .
\end{eqnarray}
Here $c$ is another numerical parameter.
The derivative is now replaced by the covariant derivative, 
because $T$ is charged under the gauge field $A_\m^{(-)}$.
We choose the following gauge choice
\begin{eqnarray}
x_8A_9-x_9A_8=0.
\end{eqnarray}
Under this gauge choice, the equations of motion can be solved by 
\begin{eqnarray}
T=q(x_8+ix_9), \quad 
A_\m^{(-)}=0 ,
\label{gaucho}
\end{eqnarray}
with $4q^4 p_1 =1$. 
Considering the $A_\m^{(-)}$ gauge fluctuations around the vortex
solution, we obtain the action
\begin{eqnarray}
S &=& {2T_{\rm  D9}} \int d^{10} x \ 
e^{-\frac{q^2}{a}(x_8^2+x_9^2)}
\Big(
q^2(x_8+x_9^2)A_{\m}^2
+(x_8^2+x_9^2)^2A_{\m}^2
+4q^4p_2(x_8^2+x_9^2)(A_8^2+A_9^2)
\no & & \hspace*{3cm}
+\frac{b}{4}F_{\m\n}^2
+2cq^2\Big[F_{\mh 8}x_8 A_{\mh}
+F_{\mh 9 } x_9 A_{\mh} 
\Big]
+(\mbox{cross terms})
\Big) .
\end{eqnarray}
After the field redefinition  $\ah_\m\equiv A_\m
\exp(-q^2(x_8^2+x_9^2)/2a)$, 
we have
\begin{eqnarray}
 S &=&
{2T_{\rm  D9}} \int d^{10} x 
\left(
\ah_{\mh}
\left[
\frac{b}{2}\left(-\p_8^2-\p_9^2+
\frac{q^4}{a^2}(x_8^2+x_9^2)-\frac{2q^2}{a}\right)
\right.
\right.
\hspace*{3cm}
\no & &\hspace*{2.5cm}
\left.
+2cq^2\left(-\frac{q^2}{a}(x_8^2+x_9^2)+1\right)
+(1+q^2)(x_8^2+x_9^2)
\right] \ah_{\mh}
\no & &
\hspace*{7.5cm}
\left.
+\frac{b}{4} \hat{F}_{\mh\nh}^2+ (\mbox{$A_8,A_9$ terms})
\right) . 
\label{amaction}
\end{eqnarray}
Since we have already used the gauge freedom, we cannot eliminate the
cross terms unfortunately. In the following analysis, we simply put 
$\hat{A}_8=\hat{A}_9=0$ for our convenience. Then the 
mass squared for the $\hat{A}_\mh$ is
\begin{eqnarray}
m_{A^{(-)}_\mh}^2=(2n+2)
\sqrt{\frac{2}{b}\left(
1+q^2+\frac{bq^4}{2a^2}
-c\frac{2q^4}{a}\right) }
-\frac{2q^2}{a}+c\frac{4q^2}{b},
\quad
n\geq 0.
\label{am-mas}
\end{eqnarray}
Whether the lowest mass squared is positive, negative or zero 
depends on the numerical value of, in particular, $c$. 
If we set $c=0$, the lowest mass squared
becomes positive, as desired.\footnote{
If we substitute the numerical values obtained in \cite{Terashima} 
into the mass formula (\ref{am-mas}), we obtain an undesirable result
in which inside of the square root in (\ref{am-mas}) becomes negative. 
The origin of this result gets back to the action (\ref{amaction}). 
The sign of the coefficient of the terms quadratic in the gauge
fields becomes negative, thus we cannot solve the Schr\"odinger
problem with the unbounded potential.
However, in the BSFT or the sigma-model approach, 
the coefficients $p_1 = 4q^4, a, c$ are ambiguous and depend on the
renormalization scheme \cite{Tseytlin1}. 
Furthermore, since we have used the derivative-truncated action,
the field redefinition does not close in our scheme.
Hence it would be less sensible 
to substitute the values of the coefficient
obtained in the BSFT. We thank S.\ Terashima for giving us
a helpful comment on this point.
}
 
We have seen that the parameters in the action are sensitive to the
resultant fluctuation spectra. In the following, we present another
approach which might be better than the approach above, in the sense
that we need no information on $c$. 

The sensitive terms are higher derivatives, thus we stick to the
two-derivative truncation. Noting that Derrick's theorem does not
mention about anything on the existence of a singular solution, we
introduce a regularization 
\begin{eqnarray}
S={2T_{\rm  D9}} \int d^{10} x \ e^{-|T|^2/a}
\Big(1+(D_\m T D^\m \bt)^{1+\e}+\frac{b}{4} F_{\m\n}^2
\Big) .
\end{eqnarray}
Here $\epsilon$ is a regularization parameter, and we take $\epsilon
\rightarrow 0$ later.
Under the gauge condition (\ref{gaucho}), we obtain a classical solution
\begin{eqnarray}
T=q(x_8+ix_9), \ A_\m=0  , 
\end{eqnarray}
where $2q^2 = \e^{-1} \to \infty$.
This solution is singular and thus avoiding Derrick's theorem.
For investigation of the fluctuation spectra, 
we redefine the gauge fields in the same manner, 
 $\ah_\m\equiv A_\m \exp(-q^2(x_8^2+x_9^2)/2a)$.
In the limit $\e \to 0$, the action for the fluctuation is given as
\begin{eqnarray}
{2T_{\rm  D9}}\! \int\! d^{10} x  \!
\left[
q^2(x_8^2\!+\!x_9^2){\ah}_\m^2\!+\!\frac{b}{2}
\left\{
\left(\frac{q^4}{a^2}(x_8^2\!+\!x_9^2)
\!-\!\frac{2q^2}{a}
\right)\!{\ah}_{\mh}^2
\!+\!\frac{q^4}{a^2}
\left(
x_8^2 {\ah}_9^2+x_9^2 {\ah}_8^2
\right)
\!+\!\frac{1}{2}\hat{F}_{\m\n}^2
\right\}
\right] 
\end{eqnarray}
where we extract up to second order of $\ah_\m$.
Since we still have the cross terms again, we simply neglect
the gauge fields $A_{8,9}$, then the mass squared for 
$A_{\mh}$ is given by\footnote{We can say
  that this result is the same as what one obtains in (\ref{am-mas}) 
with the two-derivative truncation limit $p_1\rightarrow 0$
(equivalently, $q \rightarrow \infty$) and $c=0$. In
this sense, we have consistency.}
\begin{eqnarray}
m_{A_\mh^{(-)}}^2
=(2n+2)\sqrt{\frac{q^4}{a^2}+\frac{2q^2}{b}}-\frac{2q^2}{a}, 
\quad n \geq 0.
\label{97-a-}
\end{eqnarray}
Thus the mass tower starts from the massive mode, which is
consistent with the fact that the BPS D7-brane has a single massless
gauge field coming from $A^{(+)}$ in its effective theory. 


\subsection{Universality of the fluctuation spectra}

We observed in Sec.\ \ref{gen} that, as for the kink solution, 
the results on the fluctuation in the two-derivative
truncation are universal features found in a general action. 
In this subsection, following this success, we apply this
generalization also to the vortex solution. 

The general action we consider is the same as before, (\ref{geneac}).
If we include the gauge field $A^{(-)}$ in the action, the
number of the gauge-invariant terms is quite large. Thus to avoid this
complication we again shall not consider this gauge field. 
We demand the solution to be the vortex configuration, $T=q(x_8+ix_9)$,
$A^{(+)}=0$.  Then the equation of motion becomes the following condition
for $f(X,Y)$: 
\begin{eqnarray} \label{sft-eq}
\left[f-2q^2\frac{{\de} f}{{\de} X}\right]_{X=2q^2, Y=0}=0 .
\end{eqnarray}
One of the preferable features of the generalized action (\ref{geneac})  
is that one can construct a D7-D5 bound state and (F, D7) bound 
state with correct ratio of the tensions by turning on the gauge fields
on the worldvolume of the D7-brane vortex. See App.\ A for the details. 

First, let us examine the tachyon fluctuations:
$T=q(x_8+ix_9)+t_1(x)$. The quadratic action to which the fluctuation
is subject is found as 
\begin{eqnarray}
{2T_{\rm  D9}} \frac{f(2q^2,0)}{2q^2}\!\int\! d^{10} x   
\left[(\p_\mh \h)^2
+ C_2 \h \left( -\p_8^2\!-\!\frac{q^2}{a}\!+\!
\frac{q^4}{a^2}x_8^2 \right) \h
+ C_3 \h \left( -\p_9^2\!-\!\frac{q^2}{a}\!+\!
\frac{q^4}{a^2}x_9^2 \right) \h
\right],
\nonumber
\end{eqnarray}
where
\begin{eqnarray}
  C_2\equiv \left[
1+\frac{4q^4}{f}\left(2\frac{\delta f}{\delta Y} + \frac{\delta^2
    f}{\delta X^2}\right)
\right]_{X=2q^2, Y=0},\quad
  C_3\equiv \left[
1+\frac{8q^4}{f}\frac{\delta f}{\delta Y}
\right]_{X=2q^2, Y=0}.
\end{eqnarray}
In obtaining the above action,  we have already redefined 
$\h \equiv t_1\exp(-q^2(x_8^2+x_9^2)/2a) $ and used the condition 
(\ref{sft-eq}).
The mass squared is 
\begin{eqnarray}
\label{masssqu}
m_{t_1}^2=\frac{2q^2}{a} (C_2 n_1 + C_3 n_2) 
\quad \mbox{where}\quad n_i \geq 0 , \quad i=1,2,
\end{eqnarray}
therefore the lowest mode is a massless mode.
This is consistent with the previous analysis in Sec.\ 3.1.

Next, we study the gauge fluctuation $A^{(+)}$.
The action for the fluctuation $A_\m(x)$ is given by
\begin{eqnarray}
S={2T_{\rm  D9}}\frac{bf(2q^2,0)}{4}
 \int d^{10} x \ 
e^{-\frac{q^2}{a}(x_8^2+x_9^2)} 
\left((F_{\mh\nh}^{(+)})^2-2(F_{89}^{(+)})^2\right) .
\end{eqnarray}
Here, again, interesting action is obtained (recall eq.\ 
(\ref{gaugeflu})). 
Above action tells us not only that the gauge potentials $A_{8,9}(x)$ 
are not  mixed with the rest $A_\mh$, but also that $A_{8,9}(x)$ 
have no kinetic term to run along the world volume of the
D7-brane. Therefore they completely decouple from the physics on the
D7-brane, except for the constant modes. 

Thus, we showed the universality of the results obtained in Sec.\ 3.1
and 3.2. This general analysis shows that the field content on the
vortex (= the BPS D7-brane) is properly given from the fluctuation
spectra. The decent relation concerning the tachyon condensation 
from the D9$\bar{\rm {D}}$9 system to the BPS D7-brane in Type IIB
string theory is examined at the level of the low-energy effective
field theories.  


\subsection{Derivation of the Dirac-Born-Infeld action}

Let us construct the effective action for the lowest fluctuation modes. 
The quadratic approximation gives only a trivial action of massless
scalars and a gauge field. We would like to go further in order to see
the resultant brane physics. In this subsection, we will find that the
DBI action is derived in the following manner.

Let us focus on the massless modes $t_1$ and $t_2$. 
We can give them a meaning of the displacement fields, and
this enables us to go beyond the quadratic approximation.
In this subsection
we turn on the constant gauge field strength $F_{\mh\nh}$ on the
vortex worldvolume, and treat them as a background. We follow the
procedure developed in \cite{Tseytlin2}. At last we obtain the DBI
action for the vortex D7-brane. 

First, since the fields $t_{1,2}$ represent the transverse locations of
the vortex in the $x_8$-$x_9$ plane, we can put them into the argument
of the tachyon solution:
\begin{eqnarray}
  T = T_{\rm  sol}(y_8, y_9),\quad 
y_8 \equiv \frac{1}{\beta_8} \Bigl( x_8 + t_1(x_\mh)\Bigr), 
\quad
y_9 \equiv \frac{1}{\beta_9} \Bigl( x_9 + t_2(x_\mh)\Bigr).
\label{displacement}
\end{eqnarray}
Here $\beta$'s are the boost factor obtained in the following
manner. Let us consider $t_i$ ($i=1,2$) linear in $x_\m$. 
These correspond to a special
Lorentz transformation in the bulk, that is, a rotation of the vortex
brane: $x \mapsto y \equiv \Lambda x$. 
Then, this rotation must preserve the 
metric on the brane as
\begin{eqnarray}
(\Lambda^t) G \Lambda = G.
\label{GG}
\end{eqnarray}
The metric on the brane is the open string metric
\begin{eqnarray}
G^{\mh\nh} =\left(
\frac{1}{1-bF^2}
\right)^{\mh\nh},\quad G^{88} = G^{99} = 1,
\end{eqnarray}
and $\Lambda$ is related to the displacement (\ref{displacement}) as
follows: 
\begin{eqnarray}
  \Lambda = \left(
    \begin{array}{cc|c}
1/\beta_8 & 0 & \p_\mh t_1/\beta_8\\
0 & 1/\beta_9 & \p_\mh t_2/\beta_9\\
\hline
{}* &* &* \\
\vdots &\vdots &\vdots
    \end{array}
\right).
\label{lambda}
\end{eqnarray}
We determine the unknown parameters in the Lorentz transformation
matrix $\Lambda$ by substituting (\ref{lambda}) into the condition
(\ref{GG}).
The result is
\begin{eqnarray}
  \beta_8 = \sqrt{1+ G^{\mh\nh}\p_\mh t_1 \p_\nh t_1},\quad
  \beta_9 = \sqrt{1+ G^{\mh\nh}\p_\mh t_2 \p_\nh t_2},\quad
 G^{\mh\nh}\p_\mh t_1 \p_\nh t_2=0.
\label{reb}
\end{eqnarray}
The first and the second equation determine the boost factor.
The third equation indicates that the Lorentz invariance is satisfied
only if $\p_\mh t_1$ and $\p_\mh t_2$ are orthogonal to each other.
This property is due to that our ansatz (\ref{lambda}) for the Lorentz
transformation does not include the rotation among $x_8$ and $x_9$.

Knowing the boost factor, we can proceed to obtain the effective
action for $t_{1,2}$. Substituting (\ref{displacement}) into the
action (\ref{geneac}) 
and performing the integration along the directions $x_{8,9}$ 
by changing the variables to the coordinates $y_{8,9}$, we obtain
\begin{eqnarray}
  S = T_{\rm  D7} \int d^8 x \beta_8 \beta_9 \sqrt{\det
    \left(1_{\mh\nh}+\sqrt{b}F_{\mh\nh}\right) }. 
\label{prepre}
\end{eqnarray}
Using the identity valid for arbitrary matrices \cite{Gary}
\begin{eqnarray}
  \det (M) \det \left( N + A M^{-1} A^t\right)
= \det (N) \det \left( M + A^t N^{-1} A\right)
= \det
\left(
\begin{array}{cc}
M & -A^t \\ A & N
\end{array}
\right),
\end{eqnarray}
it is possible to show that the action (\ref{prepre}) 
is in fact the DBI action of the BPS D7-brane, 
\begin{eqnarray}
  S = T_{\rm  D7} \int d^8 x \sqrt{\det
    \left(1_{\mh\nh}+\sqrt{b} F_{\mh\nh}+ \p_\mh t_1 \p_\nh t_1 
+\p_\mh t_2 \p_\nh t_2  \right) }. 
\label{dbi}
\end{eqnarray}
Here we have used the relations (\ref{reb}) and another
orthogonality $\p_\mh t_1 (FG)^{\mh\nh} \p_\nh t_2 =0$. 

Finally we have obtained the DBI action as an effective action of the
vortex solution. Therefore, at the effective action level, we have
shown that Sen's conjecture is correct, on the part concerning the
vortex-type tachyon condensation, a D9$\bar{\rm {D}}$9 pair
$\rightarrow$  a BPS D7-brane.


\section{Conclusion and Discussion}

In this paper, we confirmed a part of the descent relations in Sen's
conjecture, by using the low energy effective actions of
a brane antibrane system.  
First, in Sec.\ 2, 
we analyzed the tachyon condensation from D9$\bar{\rm {D}}$9 
system to non-BPS D8-brane using the effective action of the former.
The action (\ref{98-ac}) allows a tachyon kink solution (\ref{98-sol}),
and  
the fluctuation around them reproduces the precise low energy field
content on a non-BPS D8-brane, (\ref{98-fluta}) (\ref{98-t2}) 
(\ref{fluctuationA-}). 
The lowest fluctuation modes of the tachyon and the gauge fields obey
the action of the Minahan-Zwiebach model (\ref{effec}) 
which is an effective theory of the non-BPS D8-brane, thus we
verified the descent relation (1) in Fig.\ 1 at the level of effective
field theories. 

Second, in Sec.\ 3, 
we carried out the similar analysis for the tachyon condensation
(2) in Fig.\ 1, that is, from the D9$\bar{\rm {D}}$9 system to a BPS
D7-brane, (\ref{97-t1}) (\ref{97-a+}) (\ref{97-a-}). 
We obtained correct low energy matter content on the
D7-brane, and gave a positive evidence for that they are subject to the
Dirac-Born-Infeld action (\ref{dbi}).

Furthermore, in Sec.\ 2.3 and 3.3, 
we extended these analyses to include an arbitrary form of the action,
and obtained the same results. Namely,  a universality was found in 
the descent relations.

We should make comments about the assumptions we made in this
paper and their validity. 
Throughout the paper we studied fluctuation spectra for field by
field, however generically there are mixing terms even at the
quadratic level. It seems very difficult to diagonalize the total
system, and thus we neglected these cross terms naively, then the mass 
squared of the fluctuations are given correctly. 
One of the cross terms is between $t_1$ and $t_2$ in Sec.\ 3.1. One
can show that the lowest modes of $t_1$ and $t_2$ obtained there satisfy
also the total equations of motion of them with mixings, hence this
mixing seems irrelevant. However, we have no idea how to treat
$A^{(\pm)}_{8,9}$ and their mixing to $A^{(-)}_{\mh}$ in the vortex
analysis in Sec.\ 3.2.  The only salvation is that in the general
treatment of Sec.\ 3.3, the mixing terms for $A^{(+)}$ vanish
generically. Thus we expect that this kind of decoupling may
happen also for $A^{(-)}$ if we properly include higher terms in
$A^{(-)}$.  
 
If we turn our eyes to the future possibility of the effective field
theory approach of the D$p\bar{\rm D }p$ system and the tachyon
condensation, we find many directions. It is important to 
generalize our analysis to the non-Abelian case and check the
fluctuation spectra around lower-dimensional D-branes via
Atiyah-Bott-Shapiro (ABS) construction. It is interesting 
to construct the various brane configuration in the brane antibrane
system using the effective theory, following the construction in
\cite{hahi2,hahi1}. Some of the brane bound states are studied in the
appendix. 

The final discussion is devoted to the following question: what is the
concrete form of the effective action which reproduces all the desirable 
properties appearing in the descent relations?
As shown in Sec.\ 2.3 and 3.3, we can get universal results on the
fluctuation spectra by allowing rather general action (\ref{geneac}). 
Now, since we obtained all the mass spectra and various properties of the
defects, we can demand that these properties should reproduce the
appropriate values in string theory --- the values concerning mass
spacing, tensions and tachyon masses. Let us list our results and
see how this requirement constrains the arbitrary function $f(X,Y)$ in
the action.

\begin{itemize}
\item Potential height. The potential height should be consistent with
  Sen's conjecture: the true vacuum of the tachyons can be interpreted
  as the annihilation of the two branes. Therefore, 
\begin{eqnarray}
 f(0,0) = 1.
\end{eqnarray}


 \item Perturbative tachyon masses.
In the original system of brane antibrane, we expand the potential and
the function $f$, then the tachyon mass squared  
is given by 
\begin{eqnarray}
\left[\frac{-f}{a}\left(\frac{\delta f}{\delta
      X}\right)^{-1}\right]_{X=Y=0}  
= \frac{-1}{2\alpha'}.
\end{eqnarray}


 \item Existence of the kink and vortex solutions. This requirement 
is (\ref{sft-eq3}) and (\ref{sft-eq}), the existence of $q_k$ (for the
kink) and $q_v$ (for the vortex) which satisfy 
\begin{eqnarray}
\left[f-2q_k^2\frac{{\de} f}
{{\de} X}-4q_k^4\frac{{\de} f}{{\de} Y}
\right]_{X=q_k^2, Y=q_k^4}=0 ,\quad
    \left[f-2q_v^2\frac{{\de} f}{{\de} X}\right]_{X=2q_v^2, Y=0}=0 .
\end{eqnarray}

\item Tension of the defects.
The kink solution (a non-BPS D8-brane) and the vortex solution (a BPS
D7-brane) have their tensions ${\cal T}=\sqrt{2}T_{\rm D8}$ 
and $T_{\rm D7}$ respectively,
\begin{eqnarray}
{2T_{\rm  D9}} f(q_k^2,q_k^4)\frac{\sqrt{a\pi}}{2q_k}
=\sqrt{2}T_{\rm D8},\quad
{2T_{\rm  D9}} f(2q_v^2,0)\frac{a\pi}{4q_v^2}=T_{\rm  D7}.
\end{eqnarray}

\item Mass spectrum of the fluctuation around the kink.
From the mass spacing of $t_1$ and $t_2$, we have respectively
\begin{eqnarray}
&&\left[
 \frac{4q_k^4}{af}
\left(\frac{\delta f}{\delta X}
+6q_k^2\frac{\delta f}{\delta Y}
+2q_k^2 \frac{\delta^2 f}{\delta X^2}
+8q_k^6 \frac{\delta^2 f}{\delta Y^2}
+8q_k^4 \frac{\delta^2 f}{\delta X \delta Y}.
\right)
\right]_{X=q_k^2, Y=q_k^4}
=\frac{1}{\alpha'}, \hspace{10mm}
\\
&& \left[
\frac{f}{a} 
\left(2\frac{\delta f}{\delta X}-\frac{f}{2q_k^2}\right)^{-1}
\right]_{X=q_k^2, Y=q_k^4}
=\frac{1}{\alpha'}. 
\end{eqnarray}


\item Mass spectrum of the fluctuation around the vortex.
The spacing of the series of the mass squared is identified with the
corresponding value in string theory,
\begin{eqnarray}
\frac{2q_v^2}{a}
\left[
\left(
1+\frac{8q_v^4}{f}
\frac{\delta f}{\delta Y} 
\right)
\right]_{X=2q_v^2, Y=0} = \frac{1}{\alpha'},
\quad
\frac{2q_v^2}{a}
\left[
\frac{4q_v^4}{f}
\frac{\delta^2 f}{\delta X^2}
\right]_{X=2q_v^2, Y=0}  = \frac{n}{\alpha'},
\end{eqnarray}
for a certain positive integer $n$. Note that these two equations come
from the two spacings of the mass squared in eq.\ (\ref{masssqu}). 


\item Consistency with the Minahan-Zwiebach model.
If we simply put $T_2=0$ and replace $2T_{\rm D9}$ with 
$\sqrt{2}T_{\rm D9}$ in the action (\ref{geneac}), 
then we should have the non-BPS
D9-brane action: 
\begin{eqnarray}
 S=\sqrt{2}T_{\rm D9}\int \! d^{10}x \;
e^{-T^2/a}\sqrt{\det ( 1 + \sqrt{b} F)} \; 
f(X,Y\!=\!X^2).
\end{eqnarray}
The tachyon field is real, and thus we have a relation $Y=X^2$.
The analysis for this action was given by Minahan and
Zwiebach \cite{Zwie,MZ1,MZ2}, and that should be inherited to our
action.  Actually, the analysis for the kink in the above list gives 
sufficient conditions
for the kink analysis of Minahan and Zwiebach.

\end{itemize}

Unfortunately, we observe that
the above conditions are far less than what determines the form of the
function $f(X,Y)$. In fact, the above relations only give  particular
relations only at given values of $X$ and $Y$, thus there is no
information of functional degrees of freedom. 

However, assuming that the parameters $q_k$ and $q_v$ diverge
(which is actually realized in the BSFT), then we can extract some
information on $f$ around the infinities of the arguments. 
The result is that we can choose various parameters in $f$ 
in such a way that all of the above conditions are
satisfied.


\vspace{10mm}
\noindent
{\large \bf Acknowledgment}

K.\ H.\ would like to thank S.\ Hirano, T.\ Kuroki, T.\ Suyama, 
T.\ Takayanagi and S.\ Terashima for useful discussions and comments.


\appendix

\section{Bound States}

In this appendix, we construct various bound states of D-branes by
generalizing the vortex solution (\ref{4d-sol}) in the generalized
action (\ref{geneac}). This is a supporting evidence for the analysis
using the general action (\ref{geneac}).

In the action (\ref{geneac}), the gauge field strength for $A^{(+)}$
is coupled to the tachyon only through the metric $G^{\m\n}$ and the
Born-Infeld-type front factor. Therefore, if the following three
conditions are satisfied simultaneously, the total equations of motion
are solved: 
\begin{itemize}
\item[(i)] The tachyon configuration is (\ref{4d-sol}):
  $T=q(x_8+ix_9)$, and the function $f$ in the general action
  satisfies the condition (\ref{sft-eq}) again. 
\item[(ii)] The gauge fields transverse to the vortex worldvolume, 
$A_{8,9}$, vanish. 
\item[(iii)] The other gauge fields $A_{\mh}$ depend only on $x_0,
\cdots, x_7$, and are subject to the Born-Infeld equations of motion. 
\end{itemize}

In particular, the equations of motion for the Born-Infeld theory can
be solved of course by a constant field strength. This leads us to
brane bound states. As an example, we turn on only
$F_{67}^{(+)}=\rm {const}$. Then the resultant vortex solution is just a
D7-D5 bound state in which the uniformly distributed D5-branes have their
worldvolumes along $x_{0,1,2,\cdots,5}$. The energy of the bound
state can be calculated as
\begin{eqnarray}
E_{\rm {D7D5}}=\sqrt{\det \left(1+\sqrt{b}F\right)}
{2T_{\rm  D9}}  \frac{a}{q^2}
 f(2q^2,0)  \int d^{8} x . 
\end{eqnarray}
On the other hand, the D7-brane energy is 
\begin{eqnarray}
E_{\rm {D7}}=
{2T_{\rm  D9}}  \frac{a}{q^2}
 f(2q^2,0) \int d^{8} x . 
\end{eqnarray}
Thus we can obtain the correct result well-known in string theory,
\begin{eqnarray}
E_{\rm {D7D5}}=\sqrt{\det \left(1+\sqrt{b}F\right)}E_{\rm {D7}} .
\end{eqnarray}
This result is analogous to what was found in \cite{Tseytlin2}.
In addition, let us see how the Ramond-Raomnd charge of the
lower-dimensional branes is realized in the CS coupling \cite{KL, Terashima}
\begin{eqnarray} \label{sft-CS} 
S_{\rm {CS}} 
={2T_{\rm  D9}}  \int C \wedge {\rm Tr} \sigma_3 \exp 
 \left(\begin{array}{cc}
      F^{(1)}-|T|^2  &  d\bt \\
      dT &  F^{(2)}-|T|^2
        \end{array}\right) .
\end{eqnarray}
Here we are sloppy about the numerical coefficients.
Substituting the solution into eq.\ (\ref{sft-CS}),
we can obtain the D5-brane charge and the 
D7-brane charge:
\begin{eqnarray}
S_{\rm {CS}} 
={2T_{\rm  D9}} 
\int e^{-|T|^2/a}C^{(6)}\! \wedge\!
(F^{(+)}_{67}dx_6\! \wedge\! dx_7)\! \wedge\! dT\! \wedge\! d\bt
+{2T_{\rm  D9}} 
\int e^{-|T|^2/a}C^{(8)} \!\wedge\!
dT\! \wedge\! d\bt .
\end{eqnarray}

Another example is an (F,D7) bound state. 
Turning on an electric component $F_{07}=\rm {const}$,
the energy of the Hamiltonian density is given as
\begin{eqnarray}
{\cal H}_{\rm {(F,D7)}}={\cal L}- \Pi^\m \dot{A}_\m
=\frac{1}{\sqrt{1-bF^2}} {\cal H}_{\rm {D7}} .
\end{eqnarray}
Therefore we obtain 
\begin{eqnarray}
E_{\rm {(F,D7)}}=\frac{1}{\sqrt{1-bF^2}}
E_{\rm {D7}} ,
\end{eqnarray}
which is identical  with the well-known result in string theory.

These phenomena intrinsically occur since the Born-Infeld action is
factored out in the general action (\ref{geneac}). Therefore, we can
apply all the results in Born-Infeld dynamics in string theory,
especially the various brane configurations realized in Born-Infeld
theory.

\newcommand{\J}[4]{{\sl #1} {\bf #2} (#3) #4}
\newcommand{\andJ}[3]{{\bf #1} (#2) #3}
\newcommand{\AP}{Ann.\ Phys.\ (N.Y.)}
\newcommand{\MPL}{Mod.\ Phys.\ Lett.}
\newcommand{\NP}{Nucl.\ Phys.}
\newcommand{\PL}{Phys.\ Lett.}
\newcommand{\PR}{ Phys.\ Rev.}
\newcommand{\PRL}{Phys.\ Rev.\ Lett.}
\newcommand{\PTP}{Prog.\ Theor.\ Phys.}
\newcommand{\hep}[1]{{\tt hep-th/{#1}}}

\end{document}